\documentclass[sigconf]{acmart}
\usepackage{enumitem}

\AtBeginDocument{%
  }

\setcopyright{acmlicensed}

\copyrightyear{2026}
\acmYear{2026}
\setcopyright{rightsretained}
\acmConference[DIS '26]{Designing Interactive Systems Conference}{June 13--17, 2026}{Singapore, Singapore}
\acmBooktitle{Designing Interactive Systems Conference (DIS '26), June 13--17, 2026, Singapore, Singapore} \acmDOI{10.1145/3800645.3812959}
\acmISBN{979-8-4007-2563-0/2026/06}

\begin{document}

\title[Social and Structural Contexts of People with Visual Impairments]{Situating the Development of Conversational Artificial Intelligence in the Social and Structural Contexts of People with Visual Impairments}

\author{Jeanne Choi}
\email{jeannechoi@kaist.ac.kr}
\affiliation{%
  \institution{School of Computing, KAIST}
  \city{Daejeon}
  \country{Republic of Korea}
}

\author{Dasom Choi}
\email{dasom@nus.edu.sg}
\affiliation{
  \institution{Division of Industrial Design, National University of Singapore}
  \city{Singapore}
  \country{Singapore}
}

\author{Sejun Jeong}
\email{sejunisbest@kaist.ac.kr}
\affiliation{
  \institution{School of Computing, KAIST}
  \city{Daejeon}
  \country{Republic of Korea}
}

\author{Hwajung Hong}
\affiliation{%
  \institution{Department of Industrial Design, KAIST}
  \city{Daejeon}
  \country{Republic of Korea}
}
\email{hwajung@kaist.ac.kr}

\author{Joseph Seering}
\affiliation{%
  \institution{School of Computing, KAIST}
  \city{Daejeon}
  \country{Republic of Korea}
}
\email{seering@kaist.ac.kr}

\renewcommand{\shortauthors}{Choi et al.}

\begin{abstract}
  People with visual impairments (PVI) increasingly adopt conversational AI (CAI) in their daily practices. While much existing HCI research has focused on the technical capabilities of CAI, less has examined the societal contexts in which PVI use CAI, particularly from non-Western perspectives. We conducted a study with 14 participants with visual impairments in South Korea using an audio-based probe featuring imagined dialogues between a blind user and a future CAI. Our findings situate CAI use alongside persistent social barriers such as prejudice and restricted employment opportunities that contribute to a lack of social visibility for PVI. These societal conditions shape not only how CAI is used, but also how the potential benefits and limitations of CAI are experienced. We discuss the need for CAI design within the socio-technical realities of PVI, and conclude by discussing the importance of emphasizing social awareness and empowerment in the development of future CAI systems.
\end{abstract}



\received{January 2026}
\received[revised]{March 2026}
\received[accepted]{April 2026}


\maketitle

\section{Introduction}
The daily lives of people with visual impairments (PVI) are deeply influenced by their social and physical surroundings. These environments determine how easily they can access information, maintain their independence, and participate in society. Navigating these environments involves performing everyday tasks, which often requires mediating visual information through digital technologies including screen readers~\cite{screenreader_01} and image recognition tools~\cite{emerging24}. In recent years, conversational AI (CAI) systems such as ChatGPT have emerged as general-purpose technologies that promise new ways for PVI to access information and accomplish tasks through natural language interaction. Early research has begun to document how PVI experiment with and appropriate these systems for everyday tasks, including information retrieval~\cite{everydayuncertain}, image understanding~\cite{probing25, findmythings, microsoft24}, and creative work~\cite{kingadnin}.

While this growing body of work provides valuable insight into the functional benefits and limitations of current CAI systems for PVI, it has largely focused on task-level interactions and system capabilities. As a result, we still know little about how broader social and structural conditions shape how PVI aspire to use and adopt CAI in their lives, particularly as these systems become more embedded in everyday social environments. Prior research on emerging and assistive technologies such as personal computers, voice interfaces or camera-based systems has shown that their adoption often reproduced or even exacerbated existing social inequalities, as marginalized communities faced compounded barriers to access and fairness~\cite{contextfirst01, ames2019charisma, pradhan2018, akter2022, cook2025}. In this light, relatively little prior work has critically engaged with questions like how persistent barriers such as societal prejudice toward PVI may constrain not only access to CAI, but also the motivation and confidence to engage with future CAI systems. Without attending to these socio-technical dynamics, there is a risk that CAI reproduces existing inequalities rather than transforming them.

In this paper, we argue that understanding how PVI would use CAI requires moving beyond questions of technical assistance to examine social visibility, exploring how PVI should be engaged as autonomous social actors. 
Here, we discuss \textit{social visibility} as a state that extends beyond physical presence to the broader social standing and opportunities afforded to PVI within society. 
From this perspective, CAI is not merely a tool for completing tasks faster or more easily; it is a technology deeply embedded in social structures. Its role and meaning are shaped by structural barriers and the lived experiences of marginalization within the PVI community. Attending to these dimensions is essential for designing future CAI that meaningfully supports autonomy and inclusion of PVI.

To investigate these issues, we ask:

\begin{itemize}[topsep=1pt, itemsep=0pt, parsep=0pt, partopsep=0pt]
    \item \textbf{RQ1.} In what social contexts do people with visual impairments envision using future conversational AI?
    \item \textbf{RQ2.} What social and structural factors shape the aspirations, concerns, and perceived limits of future conversational AI for people with visual impairments?
\end{itemize}

We conducted an interview study with fourteen participants with visual impairments in South Korea, a context that has been largely underrepresented in accessibility and HCI research on PVI. While most prior studies of PVI and assistive technologies have focused on Western settings~\cite{object_02, microsoft24}, South Korea presents distinct structural conditions, including highly constrained career pathways for PVI and persistent public prejudice that shape everyday life and technology use~\cite{koreapvimass, koreapvimass02, koreapvimass03}. Rather than offering a comparative analysis, this study broadens the empirical and cultural scope of inclusive design research by foregrounding perspectives that are often absent from dominant narratives.

To support participants in reflecting on futures that are not yet technically realizable, we used a probe consisting of audio-based dialogues between an imagined future CAI system and a user with total blindness. The probe was developed through an initial formative interview phase with fourteen participants, which informed a conceptual foundation and dialogue scenarios of the probe. It was then refined through expert feedback and implemented as a fully audio-based artifact to ensure accessibility. The same fourteen participants were subsequently invited to engage with the probe in an interview study, and five participants took part in follow-up interviews to further elaborate on the societal and structural barriers surrounding CAI use. By grounding imagined futures in the lived experiences of PVI with current CAI tools, the probe enabled participants to articulate not only potential opportunities, but also the societal and psychological conditions that would shape whether such futures feel attainable.

Our findings reveal that while participants could readily imagine how future CAI could expand access to mobility and improve task performance, these possibilities were consistently constrained by social barriers. Participants described how limited career opportunities, narrow social networks, and enduring societal prejudice collectively produce a lack of social visibility that shapes how they view their own potential and autonomy. Importantly, participants emphasized that technological advancement alone does not resolve these challenges; CAI becomes entangled with existing inequalities, sometimes amplifying self-doubt or reinforcing dependence. At the same time, participants articulated nuanced aspirations for CAI as a catalyst for confidence-building and increased public presence if designed in ways that support agency by respecting the sensory realities of visual impairment, including the need for tactile interactions or heterogeneity of visual impairment.

This paper makes the following contributions.
First, we provide an empirically-grounded account of how social visibility, structural inequality, and internalized prejudice shape how PVI imagine and engage with future CAI.
Second, we translate these insights into socio-technical design considerations for CAI that foreground autonomy, confidence, and public participation, extending discussions of accessibility beyond functionality toward ethical, social, and experiential dimensions of CAI design.


\section{Related Work}
In this section, we review prior work relevant to understanding how PVI engage with CAI within broader social and structural contexts. 
We begin by situating CAI within a longer history of software technologies that PVI actively appropriate to navigate digital and physical environments. 
We then focus specifically on recent research documenting the emerging practices, benefits, and limitations of CAI use by PVI. Here, we broadly scope CAI as a software system capable of engaging in natural‐language interaction with users~\cite{cadef}. 
Finally, we review the societal barriers PVI face in daily and social life, establishing a foundation for understanding the social and structural contexts in which CAI use occurs.

\subsection{Evolving Software Technology for People with Visual Impairments}
People with visual impairments (PVI) have long engaged with a wide range of software technologies to support access and participation in everyday life. The first of these is in accessing digital interfaces through screen readers, which read on-screen content and navigation cues aloud. Built-in options in mobile phones like Apple VoiceOver~\cite{voiceover} and Android TalkBack~\cite{talkback} have made mainstream devices widely accessible~\cite{screenreader_01}. Another key domain is interpreting surroundings, which is important for obtaining essential visual explanations for daily tasks~\cite{microsoft24}. Services like VizWiz~\cite{vizwiz10} and Be My Eyes~\cite{bemyeyes} connect users with remote sighted helpers~\cite{rsa_01}, while recent AI-leveraged tools such as Sullivan Plus~\cite{sullivan} or SeeingAI~\cite{seeingai} use cameras to recognize objects, read text, and describe scenes~\cite{microsoft24, object_01, object_03}. These camera-based assistive vision systems allow users to accomplish everyday tasks such as locating personal belongings~\cite{findmythings, object_02} or reading out visual features of objects including color~\cite{microsoft24}. Various interfaces have also been designed to support PVI in exploring visual content through touch-based interaction~\cite{imageexplor21, imageassist23}. These technologies now even support navigation in physical spaces, which has long been a critical challenge for PVI,  by reading navigation cues such as signposts~\cite{emerging24, envision, navigate_01}.

Despite their promise, current assistive tools have notable limitations including accuracy limitations in object recognition~\cite{emerging24, microsoft24}, inaccessible digital interfaces~\cite{unlabeled18, inaccess_01}, and difficulty adapting to personalized contexts especially in navigation~\cite{visionneed}. Moreover, these assistive tools often work in isolation and hardly communicate or coordinate with each other, burdening users when switching between multiple aids~\cite{switching}. Here, with a more integrated system, CAI can provide a unified natural language interface that reduces fragmentation between aids in various domains~\cite{kingadnin, integratedsys} which makes their use more intuitive for PVI. The conversational modality is especially promising for PVI who may prefer interacting by voice rather than dealing with complex app interfaces~\cite{codesign_02, voice022}. Motivated by these affordances, our work focuses on CAI as a promising paradigm for understanding how PVI engage with emerging, integrated assistive technologies in practice.

\subsection{Perceived Benefits and Risks of Conversational AI Tools for People with Visual Impairment}
Conversational AI (CAI) is increasingly utilized by PVI across a variety of domains as it introduces a more general-purpose, conversational interface that blurs boundaries between assistive, informational, and creative technologies unlike many prior tools designed explicitly for accessibility. Beyond object recognition applications, PVI utilize voice assistants like Apple Siri~\cite{siri} to perform a range of daily tasks hands-free with simple speech commands~\cite{voice022, voice_12}. As those voice assistants are usually natively embedded in mobile interfaces, PVI are able to interact with apps or appliances more intuitively~\cite{siritalks18}.
More recently, LLM-based chatbots such as ChatGPT are increasingly used by PVI for information seeking, serving as more efficient and accessible knowledge companions compared to traditional search platforms~\cite{kingadnin}. 
These chatbots can also be a creative resource for PVI in domains of leisure, writing, and task management~\cite{diary24, skylimit25}. 
Moreover, PVI use CAI to handle various digital accessibility challenges including accessing digital files~\cite{uniimpact25}, or checking whether a specific website is accessible~\cite{diary24} based on the image recognition features. As voice interactions are starting to be embedded in LLM-based chatbots, this unifies the potential of each: ChatGPT has introduced an Advanced Voice feature that allows users to share videos for real-time explanations of visual content~\cite{chatgptvideo}, which integrates object recognition capabilities but also facilitates human-like dialogic interaction~\cite{probing25}. Such advancements hold potential value for PVI by providing nuanced, real-time descriptions of visual contexts~\cite{probing25, envision, skylimit25}.

Collectively, these practices illustrate the ways how PVI are starting to actively experiment various uses of CAI.
However, PVI encounter a variety of challenges when utilizing CAI in daily life, beginning with accessibility barriers such as incompatibility with screen readers~\cite{kingadnin}. In information retrieval, they must navigate inaccurate or low-quality responses, often developing their own strategies and tolerances for managing such uncertainty~\cite{everydayuncertain}. Also, due to sycophantic tendencies of such systems, PVI actively have to adapt to catch any erroneous agreements of the system in order to elicit more reliable descriptions on visual content by the system~\cite{probing25}. More broadly, PVI report that CAI does not yet adequately understand the PVI community, as outputs frequently reflect biased assumptions rooted in sighted perspectives, suggesting training data that insufficiently represents PVI experiences~\cite{probing25, kingadnin, biasedai_01}.

While this body of work provides important insight into how CAI is currently used and experienced, it largely focuses on functional interactions and system-level behaviors. Prior research on related assistive technologies has examined broader social and ethical considerations such as privacy and social stigma in the use of voice assistants and camera-based systems~\cite{pradhan2018, akter2022}. However, how these concerns manifest in the context of increasingly general-purpose and conversational CAI systems remains underexplored. Less attention has been paid to how PVI evaluate CAI in relation to broader life aspirations, confidence, identity, and participation in social life. Although speculative and futures-oriented HCI research has explored possible trajectories of CAI with various user groups~\cite{dfadolescent, dfolder, fictfail25, dynamic24, lurialetter}, these studies often foreground technological possibilities and risks rather than the lived social contexts in which such futures would unfold. Moreover, these futures of CAI are mostly imagined from sighted perspectives, leaving the aspirations and concerns of PVI underexamined. To fill in this gap, we explore the interrelated social and structural circumstances that PVI would face when using CAI.

\subsection{Societal and Structural Barriers of PVI}
Prior research in disability studies and sociology has documented the societal barriers that PVI face in both institutional settings and everyday social interactions. Research consistently shows that PVI are subject to social invisibility, where their voices and contributions are marginalized in public discourse~\cite{ableism24}. Visual impairment is often framed through deficit-oriented narratives that associate disability with diminished social and economic agency~\cite{ableism01, ableism02}. These assumptions manifest in everyday experiences such as unsolicited assistance~\cite{unsolassis}, exclusion from decision-making~\cite{charlton1998nothing}, and lowered expectations in education and employment~\cite{roulstone2005working}.

Employment and career development emerge as critical domains where the social invisibility of PVI becomes particularly pronounced. In South Korea, the context of our study, career opportunities for PVI remain narrowly confined despite policy-level commitments to disability employment quotas~\cite{koreapvimass, koreapvimass02}. Many PVI are funneled into a limited set of socially acceptable jobs, such as massage therapy or call-center work, reflecting long-standing institutional and cultural assumptions about suitable labor for PVI. Scholars have argued that these constrained career pathways not only restrict economic independence but also shape how PVI are perceived socially~\cite{roulstone2005working}. Occupational segregation reinforces stereotypes that equate visual impairment with limited professional competence, thereby reducing the visibility of PVI in diverse professional roles.

Collectively, prior work suggests that the social barriers that PVI face are deeply entangled with societal attitudes, institutional practices, and opportunities for participation. While recent work has discussed the societal barriers that PVI face, fewer studies have explicitly examined how those social conditions intersect with the emerging CAI systems. In this work, we foreground the voices of PVI by situating their experiences within the specific social and cultural context of South Korea, and in doing so, extend accessibility research beyond questions of functionality toward critical issues of lived agency and participation.

\section{Method}
In order to investigate how PVI reflect on the social dimensions of the future experiences with CAI, we conducted a user study based using audio-based probes with fourteen participants who have a range of visual impairments. The probes were designed to depict an envisioned experience for PVI in a prospective society in which a user with total blindness engages with a futuristic CAI. This study's procedure was approved through our institution's Institutional Review Board~(IRB).

\subsection{Participants}
\begin{table*}[]
\centering


\sffamily
\footnotesize
\def\arraystretch{1.3}
\begin{tabular}{cccccclc}
\toprule
\#  & \textbf{Age}   & \textbf{Gender} & \textbf{B/LV}  & \textbf{Onset}    & \textbf{Occupation}                & \textbf{CAI Tools Used}            & \textbf{CAI Usage}      \\ \toprule
\textbf{P1}  & 30-39 & F      & LV    & Acquired & Massage Therapist         & ChatGPT, GiGAGenie        & 2-3 times/week \\
\textbf{P2}  & 30-39 & M      & Blind & Acquired & Massage Therapist         & ChatGPT, Bixby, GiGAGenie & 2-3 times/week \\
\textbf{P3}  & 50-59 & F      & Blind & Acquired & Showdown Athlete          & ChatGPT                   & 4+ times/week  \\
\textbf{P4}  & 30-39 & M      & LV    & Birth    & Digital Literacy Educator & ChatGPT, Gemini, Copilot  & 1-2 times/week \\
\textbf{P5}  & 60+   & F      & Blind & Acquired & Massage Therapist         & ChatGPT, Gemini, Bixby    & 5 times/week   \\
\textbf{P6}  & 60+   & M      & Blind & Acquired & Retired                   & ChatGPT                   & 2-3 times/week \\
\textbf{P7}  & 60+   & M      & Blind & Acquired & Massage Therapist         & ChatGPT, Gemini           & 8-9 times/week \\
\textbf{P8}  & 60+   & M      & Blind & Acquired & Retired                   & ChatGPT                   & 10+ times/week \\
\textbf{P9}  & 60+   & M      & Blind & Acquired & Pastor                    & ChatGPT                   & 1-2 times/week \\
\textbf{P10} & 60+   & F      & LV    & Acquired & Massage Therapist         & ChatGPT                   & 2-3 times/week \\
\textbf{P11} & 30-39 & F      & Blind & Birth    & Remote Office Job         & ChatGPT, Bixby, Siri      & 2-3 hours/day  \\
\textbf{P12} & 50-59 & M      & Blind & Acquired & Showdown Athlete          & ChatGPT, Gemini, Adot     & 2-3 times/week \\
\textbf{P13} & 30-39 & F      & Blind & Birth    & Massage Therapist         & ChatGPT, Bixby            & 2-3 times/day  \\
\textbf{P14} & 30-39 & M      & Blind & Birth    & Massage Therapist         & ChatGPT                   & 3 times/week   \\ \bottomrule
\end{tabular}

\caption{Demographic information of the user study participants. Fourteen participants were recruited to cover a diverse range of age, gender, types of visual impairments (blind or low-vision), occupations, and onset types (from birth or acquired), as well as experience with CAI tools.}
\Description{A table with eight columns and fifteen rows. The header row contains the following columns: Identifier, Age, Gender, B/LV (Blind or Low-Vision), Onset (whether the visual impairment was from birth or acquired), Occupation, Conversational AI tools used, and CAI usage frequency. The following rows list each participant's demographic information.}
\label{tab:participant}
\end{table*}

Fourteen participants with visual impairments were recruited for the study, including ten participants with acquired vision loss and four participants with vision loss from birth. The participants were recruited at a local welfare center specialized for PVI,\footnote{In South Korea, a number of communities have welfare centers that support rehabilitation and socialization for people with visual impairments. At the time of writing, there were approximately 15 such welfare centers in South Korea according to the Ministry of Health and Welfare~\cite{welfarecount}, and the research was conducted at a welfare center in the city where this research took place.} with recruitment criteria specifying participants who have visual impairment, either low vision or total blindness, an age of over 18, and prior experience with CAI (Table~\ref{tab:participant}). Among the participants, seven were female and seven were male.

Three experts were also recruited for expert interviews intended to support design of and iteration on the probe. The recruitment for expert interviews was performed through direct e-mails to experts in the domains of CAI technology and welfare on PVI. To contextualize the experiences of the participants within institutional and technological landscapes, the recruited experts were: a rehabilitation team leader at a welfare center specialized for PVI, a CAI educator, and a CAI industry professional. 

In order to further iterate on the probe to ensure an engaging narrative flow, a novelist was also recruited through a direct email. The four experts were compensated with 25,000 KRW (approx. USD 17) for a 1-hour interview.

\subsection{Probe Design}
In our study, we used an audio-based probe featuring an imagined conversation set 10 years in the future between a CAI and a user with total blindness. The 10 year horizon was chosen as it is sufficiently forward-looking to encourage consideration of technological advances, yet not so distant as to make the probe implausible or difficult for participants to imagine. We used a fully audio-based probe, as audio-based artifacts have been shown to be accessible and effective in eliciting insights in previous design research with PVI~\cite{speaking23, remembersound24}. The probe was created by converting a text script to audio using Naver Clova Dubbing~\cite{clovadubbing}, which is a text-to-speech service specialized in Korean language. In order to create an immersive experience for participants, we selected voices with natural pronunciation and assigned them to the CAI and the blind user in the probe. Based on these design choices, the following section details the procedure through which the probe was developed and iterated.

\subsubsection{Formative Interviews}
To ground the construction of the probe in lived experiences of PVI, we first conducted semi-structured, formative interviews with the recruited fourteen participants and experts in the field. We analyzed the interview data using Affinity Diagramming~\cite{affdia}, through which we identified six overarching categories comprising a total of 84 subthemes. All formative interviews were conducted via Zoom, and lasted for 33.2 minutes on average. The interview protocol is provided in Appendix~\ref{Format PVI}, and a detailed account of the formative interview findings can be found in Appendix~\ref{Formative_Interview_Result}. In this section, we present only those findings that directly informed the probe design.

In the formative interviews with PVI, we explored their current experiences with CAI, challenges in daily life, and aspirations on how CAI might be used in their futures. Their current use of existing CAI tools were focused on daily routines such as information search or image description. Based on this experience, participants expected utilizing future CAI in ongoing challenges such as navigation or managing social life. However, participants also pointed out the importance of ethical, careful use of such technologies including acknowledging the risk of over-reliance.

In the formative interviews with experts, they pointed out the persistent social and structural barriers that PVI face in South Korea, such as limited opportunities in employment, accessibility challenges in modern infrastructures, or lack of funding and welfare programs from the government. Also, while there are emerging tools that support the solo navigation of PVI, all experts pointed out the persistent problems of current technologies such as accuracy in navigating at the meter-level. Experts were intentionally recruited to provide a system-level perspective on these societal and structural conditions, complementing formative interviews with PVI that focused more on lived experiences of everyday life and engagement with current CAI tools.

\subsubsection{Construction of an Imagined Society}
Drawing on insights from both formative participant interviews and expert interviews, we iteratively refined a vision of a future society that served as the conceptual foundation for our probe. To guide this process, we referred to the first three steps of Schwartz’s well-established eight-step scenario development framework~\cite{schwartz1997art}, which provides a structured approach to constructing the core components of a scenario environment. This process envisioned a future society characterized by markedly improved societal attitudes toward PVI and substantial technological advancements that enhanced accessibility across both digital systems and the physical environment. In the envisioned society, these developments facilitated greater mobility and information access of PVI, which in turn contributed to increased self-esteem and broader social participation. The full building process and description of the society is shown in Appendix~\ref{Future Society}.

\subsubsection{Dialogue Iteration}
Within this societal context, the specific conversation dialogues that formed the probe were developed iteratively among the research team. We discussed the most critical and prevalent issues in PVI's daily and social life that were mentioned in the formative interviews: first, independent navigation, which all participants identified as both a current challenge and a primary expectation for future CAI use; second, the discrepancy between PVI's will to participate in society and the significant social barriers shaping limited career path of PVI in South Korea.

Based on these sources, we brainstormed how an advanced CAI could be integrated in such situations. For the first dialogue on independent navigation, we selected the concept of `solo trip abroad' as it is the most challenging navigation task compared to navigating to places familiar in daily life such as restaurants, creating a futuristic, unexpected situation. For the second dialogue focused on enhanced career paths, we chose policy planner among various options (e.g., digital marketing planner, interpreter) as it could show active discussion with CAI on policies and involve public advocacy. 

We then drafted both dialogues based on both the key benefits and risks of CAI and their current challenges mentioned in formative interviews, balancing both hopeful and critical elements suggested indirectly. For example, both dialogues intentionally incorporated moments where CAI subtly encouraged dependency, such as asserting, ``It will be better with me than your friend.'' to illustrate potential risks of over-reliance, a concern commonly articulated by participants. 

Feedback from experts and a novelist further improved the dialogues’ realism and accessibility. Experts with knowledge about navigation for PVI iterated on how the futuristic CAI should interact with the blind user to correctly lead the user. Feedback from a novelist improved the narrative depth of the dialogues by adding more emotional words and phrases, making them more engaging and thought-provoking.

\subsubsection{Resulting Probes}
As a result of this iteration, two dialogue probes were developed: the (1) \textbf{Traveling Future} and the (2) \textbf{Policy Planning Future}. 

The \textbf{Traveling Future} dialogue depicts a user visiting the Musée de l'Orangerie in Paris accompanied solely by the CAI, illustrating how the user engages with the CAI when entering the museum, anticipating specific artworks, and exiting the site. This dialogue was designed to foreground issues of independent navigation. The length of the Travel Future dialogue was 4 minutes and 6 seconds.

The \textbf{Policy Planning Future} dialogue portrays a user employed as a policy planner preparing a public presentation on a proposed welfare policy for adolescents with visual impairments, depicting interactions with the CAI during the preparation phase, the delivery of the presentation, and the subsequent reflection on the event. This dialogue was constructed to highlight challenges surrounding employment opportunities of PVI. The length of the Policy Planning Future dialogue was 2 minutes and 46 seconds.

In both dialogues, the focal user was represented as being totally blind, in order to model more sustained interaction with the CAI during tasks such as spatial navigation and the creation of visually oriented content. A prompt was added at the end of each dialogue in the form of a question asked by the future CAI to the user in the dialogue, which participants were requested to answer during the interview. This was intended to provoke reflection by creating a more immersive experience in the dialogue, helping participants imagine themselves in the future environment of the dialogue. The full text for each dialogue is shown in Appendix~\ref{Future probe}.

The dialogues were personalized for each participant by (1) matching the gender of the blind user's voice in the dialogue, and (2) replacing the name that the CAI calls the user with each participant's name, in order to provide a more immersive experience.

\subsection{Study Procedure}
The same participants who took part in the formative interviews were invited to participate in an interview study based on the probe. A four week interval passed between the formative interviews and the main interview study, which was necessary for the development of the probe. To enhance the sense of presence and provide an immersive experience, the probe was played live through a laptop in a controlled, closed-room setting that minimized external noise and distractions. While this setup may not fully reflect real-world use contexts such as navigation in public environments with substantial ambient noise, it allowed participants to focus on the probe content and ensured consistency across sessions.

The study began with a paragraph describing the futuristic society, which was narrated to participants by converting a text script to audio also using Naver Clova Dubbing, helping them understand the societal background of where the probe took place. Next, participants heard each of the dialogues in order. Each was followed by a set of questions, starting with an activity where participants directly replied to the prompt given by the imagined future CAI at the end of the dialogue. In this replying activity, participants were asked to respond by considering how they would answer if they were actually in that situation. The presentation order of the dialogues was randomized. 


After replying, the participants were asked about the overall experience of the probe, starting with general questions such as parts that they most liked or disliked. They were next asked a set of more detailed questions such as their envisioned life with future CAI and any expected potential risks or gaps regarding to the intertwined societal contexts. To help expand their imagination beyond the content in the probe, they were then asked to imagine a more advanced technology and society in year 2050, and how they might imagine the experience with CAI in that future environment. The full interview protocol is shown in Appendix~\ref{maininterview}. The participants were compensated 60,000KRW in total (approx. USD 42) for participating in both the formative interviews and the interview study with the probe. The interview study with the probe took 43.4 minutes on average. Additionally, all 14 participants were invited to a follow-up interview, of whom five agreed to participate. These interviews focused on societal and structural barriers, which emerged as a salient theme during the initial probe sessions and warranted further in-depth exploration. Participants were compensated 20,000 KRW (approx. USD 14), and the follow-up interviews took 32.5 minutes on average.

\subsection{Data Analysis}
The interviews were analyzed using an approach of thematic coding~\cite{braun2006using}. The interviews were transcribed by using Naver Clova Note\footnote{https://clovanote.naver.com/}, a Korean transcription service. 
The first author conducted an initial open-coding step on the interview transcriptions using Atlas.ti,\footnote{https://atlasti.com/} identifying emergent concepts and patterns in participants' responses. The second author then independently reviewed the initial codebook and corresponding interview segments to validate the coding scheme and identify any overlooked themes. Through iterative refinement and collaborative discussion sessions, the two authors resolved discrepancies through discussion and re-examination of the original transcripts until consensus was reached and consolidated the codes. We developed a final codebook compromising 156 codes with clear definitions and exemplar quotes for each code.

Following the coding process, the first and second authors conducted collaborative affinity diagramming sessions to synthesize higher-order themes. The codes were grouped based on semantic relationships and conceptual similarities, with particular attention to how themes related to the societal contexts of CAI use. Through iterative clustering and negotiation, 11 major categories emerged. These categories include the current needs and challenges in daily and social life as PVI, desired societal changes to better support the life of PVI, and envisioned roles of CAI in supporting their daily activities and social participation.

\subsection{Positionality Statement}
All members of the research team are based in South Korea and have either lived in Korea for several years or since birth, which informed our understanding of the sociocultural context that participants described their experiences of societal and structural barriers as PVI in South Korea. The team’s background in HCI and CSCW shaped our focus on technology-mediated experiences and the role of online and institutional support systems in participants’ lives.

Several team members have prior experience working with people with disabilities, including individuals with visual impairments, which informed our efforts to design inclusive study materials and sensitized us to participants’ lived experiences during data collection and analysis. In addition, some researchers had existing relationships with a local welfare center specialized for people with visual impairments, which facilitated participant recruitment.

At the same time, none of the researchers identify as persons with visual impairments. We acknowledge that our interpretations are shaped by our perspectives and experiences, which influenced how we conducted interviews and interpreted participant narratives.

\section{Findings}
In this section, we detail findings regarding participants' reflections on the societal contexts that shape the use of CAI for PVI in response to the probe, specifically reflecting on their social and cultural experiences as PVI living in South Korea. We start by analyzing the prevailing social and structural barriers that participants identified as constraints on future CAI adoption among PVI. Then, we explore how participants' perspectives on how these barriers might be and ought to be addressed, with particular emphasis on enhancing social visibility and fostering autonomy through future CAI design.



\subsection{Existing Social Barriers Constrain Future Conversational AI Use of People with Visual Impairment}
We first examine major societal barriers mentioned by participants that could shape how PVI envision and experience future CAI, focusing on the persistent lack of social visibility faced by the PVI community. In detail, we show how structural constraints, including limited career opportunities, narrow social networks, and persistent discrimination collectively challenge the ways in which how future CAI can be imagined and adopted in everyday life for PVI. 



\subsubsection{Structurally Limited Career Paths and Economic Constraints}
\begin{quote}
    \textit{Other than being a massage therapist, I could never imagine what kind of job I might have. (P10)}
\end{quote}

For PVI living in South Korea~---~and especially for those who are blind~---~career paths and options are very limited~\cite{koreapvimass, pvijob}. Participants mentioned that by far the most common profession for PVI in Korea is massage therapy. Although there are specialized welfare centers for PVI in South Korea that provide rehabilitation programs and vocational education, there is a significant lack of diversity in vocational education programs for PVI~\cite{koreapvimass02}, which are primarily focused on training PVI to become massage therapists~\cite{koreapvimass, koreapvimass02}. Even for individuals who acquired vision loss after having been previously employed, participants reported that they were often compelled to transition into massage therapy.



Such experiences led some participants to express surprise at the prospect of PVI pursuing occupations outside of massage therapy through collaboration with CAI, as depicted in the Policy Planning Future (P1, P10), and often fueled pessimism about the future. For example, P1 had barely imagined alternative occupations, as he would ``\textit{never let myself expect more}''. These occupational constraints often led participants to develop forms of self-discouragement regarding what they perceived themselves capable of accomplishing. Such internalized limitations hindered not only their willingness to engage with new technologies such as CAI, but also their motivation to participate more broadly in society or to envision themselves undertaking new roles or opportunities. Participants articulated this concern by reflecting on a widespread sense of resignation among PVI working as massage therapists:


\begin{quote}
    \textit{What I find most unfortunate is that, from my perspective, many people working in massage therapy seem to have completely given up and invested everything into that one occupation, abandoning all other possibilities. Even if someone became a massage therapist because of visual impairment, I think it would be beneficial to shift one’s mindset to pursue general cultural experiences, attend concerts, and make meaningful use of leisure time. (P9)}
\end{quote}

Limited career prospects naturally led to economic challenges for PVI. In particular, several participants discussed how many of PVI rely on Korea's basic livelihood security benefits, receiving a fixed amount of financial assistance each month. This form of social support was described as both essential and constraining, as it discouraged efforts to pursue alternative forms of employment or self-development. Some participants even referred to themselves, somewhat ironically, as ``level-10 public officials'', drawing a metaphor to Korea’s hierarchical civil service system, where ranks begin at level 9 and increase with seniority and compensation. This metaphor likens their fixed income to that of salaried officials, while underscoring that their benefits fall below even entry-level pay, reinforcing a sense of economic stagnation.



This economic challenge is also directly linked to the ways PVI use current CAI: in order to access specialized or advanced CAI features, which can better support their particular needs, PVI often need to subscribe to upgraded plans. For example, P2 wanted to use a CAI screen-sharing feature to obtain image descriptions for website screenshots that are otherwise inaccessible but, after prolonged consideration, ultimately chose not to subscribe due to the cost. Participants explained that although current and future CAI could improve the accessibility of some parts of the world around them, the accessibility of the CAI themselves needs to be improved.

\begin{quote}
    \textit{They should make it affordable so that all of PVI can use it. People always talk about inequality. I hope it comes out at a cheap price so everyone can use it without worry, without those inequalities. (P7)}
\end{quote}

These concerns highlight how existing economic disparities and limited employment opportunities for PVI in South Korea intersect with emerging CAI technologies. Without addressing affordability and structural economic barriers, CAI risks reproducing the same inequalities that already constrain the career trajectories and everyday mobility of PVI.

\subsubsection{Narrow Social Networks and Self-Reinforcing Limits of Experience}
Another major barrier identified by participants was societal conditions that restrict PVI's access to diverse experiences and opportunities for engaging with new technologies such as CAI. In particular, participants described a perceived tendency for PVI to remain embedded within relatively homogeneous social networks, which they claimed were often centered around specialized welfare centers for PVI in South Korea. While these provide essential peer support, they may also limit exposure to diverse perspectives and alternative ways of engaging with the world. As P1 noted, ``\textit{The friends of PVI are almost exclusively other PVI. Besides family members or relatives, they hardly know anyone who isn't visually impaired.}''

Narrow social networks among participants resulted from different causes depending on whether visual impairment was congenital or acquired. Those born with visual impairment often attributed restricted social experiences to early educational and family environments, such as segregated schooling and overprotective parenting. These restrictions limited opportunities to develop diverse communication skills and engage with broader society, sometimes requiring later efforts to overcome these constraints such as enrolling in a speech class. In contrast, participants with acquired vision loss initially maintained relationships with sighted friends but gradually gravitated toward PVI communities, where shared embodied experiences including frequent challenges in everyday life (e.g., accidentally spilling a cup of water while moving through a space) reduced social pressure and emotional strain.

However, these limited opportunities for exchange with various communities often led not only to reinforced social isolation but, as some noted, also fostered \textit{self-prejudice}, which is defined as a tendency to underestimate their own potential or to see the world only within the narrow scope of their experiences. This internalized constraint shaped how participants evaluated possible activities and opportunities, such as traveling in the Traveling Future, often leading them to dismiss unfamiliar experiences as irrelevant or unattainable (P7, P12).

\begin{quote}
    \textit{From living alongside PVI and experiencing the challenges myself, I’ve noticed that their lives can feel very narrow, mostly because many haven’t had the chance to fully participate in social life. (...) PVI often carry heavy prejudices too, toward themselves and toward others. Too often, there’s a sense of ``what I know is all there is'', simply because their access to information has been so limited. (P9)}
\end{quote}

Participants described how this limited social exposure can constrain not only everyday interactions but also their willingness to explore new forms of experiences. Participants mentioned how some PVI don't feel the need to expand their abilities or learn new knowledge, instead sticking to services or technologies they originally know such as feature phones (P5). For instance, participants reported that although some individuals with visual impairments acknowledge the potential utility of CAI in everyday contexts, many nonetheless refrain from adopting it and instead continue to rely on familiar platforms such as YouTube (P6). This observation highlights a vicious cycle in which limited societal interaction narrows worldviews, reducing motivation to seek new experiences and information. This reduced motivation subsequently constrains participants’ experiences with emerging technologies such as CAI, as unfamiliar or novel use cases are less likely to be explored or perceived as relevant.




\subsubsection{Persistent Societal Prejudice}
Many participants discussed how the discriminative societal attitude and prejudice towards PVI is still pervasive in South Korea, which also reinforced their reluctance to pursue a broader range of experiences. In particular, participants noted that members of the sighted community often perceive PVI as incapable of independently conducting any daily tasks, which in turn can deprive PVI of the motivation to learn new skills or tools such as CAI.

\begin{quote}
    \textit{I sometimes hear people say, ``I could explain it to you, but since you can’t see, you can’t do it anyway.'' Hearing that really hurts deeply. Because if someone actually explained it, of course I could understand; it’s just that I can’t see. (P2)}
\end{quote}

This perception also extended to participants’ expectations surrounding CAI use. Several participants anticipated that active engagement with advanced technologies might be met with surprise or skepticism from sighted individuals, such as reactions like, ``\textit{Why would visually impaired people attempt something so complex?}'' Participants expressed that such anticipated judgments could discourage them from exploring or adopting CAI tools.

Participants especially emphasized experiences of discrimination encountered during independent navigation in public spaces. P11 noted that, even if she could independently travel as in the Traveling Future, she might still face the same stigmatizing remarks she had previously encountered when walking alone like ``\textit{Why does a visually impaired person even come outside?}'' Some participants also mentioned how a sense of pity by sighted pedestrians was often triggered simply by being seen navigating independently in public spaces. Thus, while CAI may create new opportunities for independent participation in everyday activities, these persistent societal attitudes made participants hesitate to envision a future where they independently conduct such tasks in public. 

Beyond overt discrimination, participants also identified more subtle forms of misunderstanding that pervade everyday interactions. P14 explained how fundamental differences in PVI's perception towards and interaction with the world are often reacted with incomprehension rather than accommodation: ``\textit{PVI, especially those who are congenitally blind, try to feel objects with their hands because they've never seen things before. But those who don't know well about PVI say things like ``They always have to try touching that.''}'' This kind of persistent societal prejudice diminishes PVI’s willingness to actively engage with new technologies. 

P4, a digital literacy educator at a welfare center for PVI, further observed that many learners using smartphones and CAI tools experienced anxiety and fear of making mistakes, particularly with voice-based interactions. While voice interfaces are often considered more immediately accessible for PVI~\cite{codesign_02}, P4 described how some PVI perceive them as interactionally demanding, requiring users to formulate precise verbal commands in real time, which made learners more concerned about making mistakes. As a result, learners often spoke overly cautiously or indirectly to the system, reflecting a lack of confidence in both the technology and their own ability to use it effectively. Consequently, both explicit discrimination and more subtle misunderstandings would continue to constitute significant barriers to the public and active use of CAI by PVI, limiting the realization of its promised benefits.

\subsection{Shaping Social Visibility and Autonomy in Future Conversational AI Design}
Participants consistently emphasized that while CAI holds potential to support independent task performance, its meaningful adoption is deeply intertwined with broader social conditions, including persistent prejudice or societal misconceptions about disability. Drawing on these reflections, we highlight in this section how confidence and self-efficacy are cultivated through opportunities for self-discovery and sustained engagement with society, within which technologies such as CAI may play a supportive and mediating role. At the same time, we point to the necessity of supportive social structures and careful design considerations of CAI that respect PVI’s agency and lived experiences.

\subsubsection{Cultivating Confidence and Self-Efficacy}
Due to social barriers and limited opportunities for social participation, participants identified a lack of self-confidence as one critical challenge that makes the potential experiences of technology including CAI tools ineffective if users remain reluctant to engage with new experiences. In this regard, participants emphasized the importance of fostering a mindset among PVI that they are capable in trying or starting new activities. P8 articulated that PVI should identify distinctive capabilities such as heightened auditory perception or spatial reasoning cultivated through non-visual navigation, framing visual impairment not as a limitation to overcome, but as one aspect of a fuller identity that encompasses unique strengths and capabilities. 

This suggests that empowerment could stem from how PVI choose to use these new technologies to discover and develop \textit{their own capabilities}, rather than from offloading current tasks on to CAI. Notably, participants did not view the development of such confidence as separate from technology, but they rather perceived building confidence as something that could be cultivated in conjunction with technology. For instance, P2 described how using CAI could reduce reliance on sighted assistance and instead enable more autonomous task completion:

\begin{quote}
    \textit{Normally, I would have to ask a sighted person for help. But if I use CAI and it says things like, ``Move it a few centimeters down,'' or ``Lift it slightly higher,'' and it explains things in detail, I started thinking that ``Maybe even though I won’t ever be exactly like a non-visually impaired person, wouldn’t this give me some confidence someday?'' (P2)}
\end{quote}

In this sense, CAI becomes meaningful not because it performs tasks for PVI, but because it creates opportunities for PVI to perform tasks themselves. However, some participants did not perceive confidence gained through collaboration with CAI as addressing the more fundamental dimensions of empowerment, and instead regarded it as only a limited component of cultivating deeper, internalized confidence related to one’s sense of self. From a long-term perspective, they argued that meaningful PVI–CAI collaboration should prioritize forms of confidence grounded in self-identity, personal values, and inner resilience.

\begin{quote}
    \textit{Developing the potential within oneself and building real confidence is something different. Confidence that comes from CAI helping with a few tasks is only temporary. What matters is the inner self and that is what needs to be cultivated. On the surface, you may occasionally use CAI to do things and overcome frustrations, but what allows you to endure and overcome those moments comes from your inner self, not from the fact that CAI helped you do something in a particular situation. (P9)}
\end{quote}

Extending this concern, some participants noted that collaborating with CAI could reduce self-efficacy or increase frustration in certain cases. They worried that task completion might be attributed more to following CAI’s instructions than to one’s own problem-solving strategies, extending foundational work on self-efficacy~\cite{bandura1977self} by highlighting how reliance on CAI may undermine individuals’ perceived agency. As a result, while CAI may lower barriers to attempting new tasks, participants highlighted that cultivating durable, internalized confidence remains a distinct challenge that cannot be addressed through technological support alone.




\subsubsection{Broadening Perspectives Through Knowledge Access}
Participants expressed a desire to address the constraints arising from their limited social networks and to increase their visibility within society as a means of fostering greater inclusion. While the limited visibility of PVI in society is significantly influenced by the highly restricted career opportunities available to PVI, participants suggested that the effort could begin with PVI actively seeking to broaden their knowledge of the world and to explore new methods of engagement. 

Several participants suggested that exposure to a broader range of knowledge could foster motivation to expand their perspectives, potentially challenging narrow viewpoints that PVI may hold. In particular, they anticipated that improved access to information through the use of CAI compared to previous knowledge bases such as internet search engines could support PVI in acquiring broader and more diverse knowledge. Although participants acknowledged that this does not constitute a comprehensive solution, they viewed accessing new information through CAI as a potential catalyst for fostering curiosity about the world and spark initial interest.

\begin{quote}
    \textit{It’s like driving at night on a country road. With low beams, you only see a short distance ahead. But when you switch to high beams, suddenly the whole road opens up clearly before you. Knowledge works the same way. The more you know, the further you can see and the more you can share. That’s why I hope CAI can broaden horizons for PVI. I hope it helps break stereotypes, open up perspectives, and create space for richer conversations. (P9)}
\end{quote}

Extending this notion of broadened knowledge access, several participants envisioned CAI functioning as a secretary, not only by serving as an interactive knowledge base but also by mediating contextual information that is otherwise visually inaccessible. For instance, participants described how CAI could support their understanding of situational cues in professional environments, such as recognizing when a customer approaches. By augmenting both informational access and contextual awareness, participants anticipated that such capabilities could enable them to engage more fully in work-related tasks that were previously difficult to access, thereby expanding their opportunities for social participation and contribution.
 
\subsubsection{Increasing Public Visibility and Mutual Understanding}
Participants also highlighted the need to address the stereotypes and constrained views imposed by society arising from PVI’s limited social engagement, which often undermine their motivation to explore and adopt new tools or skills, including CAI. For addressing this, they emphasized that communication with sighted individuals can help reduce misconceptions and foster mutual understanding across both communities. However, they also cautioned that poorly structured or unreflective interactions may cause psychological harm and reinforce misunderstandings. As P9 noted, differences in access to visual information can lead PVI to interpret social situations differently, yet sighted individuals may focus on outcomes rather than the underlying cognitive and experiential processes. Thus, participants stressed that meaningful cross-community communication requires sighted individuals to better understand how PVI perceive and navigate their environments. Here, some participants mentioned how CAI could contribute on this as a mediator by offering sighted people more accurate information about PVI (P1, P3, P6). For example, as PVI have different strategies to understand the world such as utilizing tactile interactions, P14 hoped that CAI could more accurately explain the unique characteristics and needs of PVI during shared interactions with sighted people.

Beyond interactions between communities, participants pointed on the importance of explicitly increasing the public visibility of PVI conducting various tasks. P11 suggested that making the independence of PVI more visible, for example, by showing how they navigate public spaces autonomously, could contribute on letting sighted people naturally develop curiosity towards PVI and more accurately recognize their agency and capabilities.
Participants also saw the importance of going beyond passive media representation to create active engagement opportunities through public demonstrations and exhibitions. 
\begin{quote}
    \textit{I think we need to inform people a lot more about these things, through education or other means. (...) When people see PVI using CAI and ask ``Why are you using CAI?'' and I can respond ``I'm currently getting help from CAI, so I'm fine,'' and there will be people who become interested and ask ``Oh, you can get that kind of help?''. When new products come out, there are exhibitions and demonstrations to let people come see what's available. I think the society should do more of that, not just showing assistive technologies on TV channels, but also raising awareness about people with disabilities and helping perceptions change more broadly. (P11)}
\end{quote}

Consequently, participants stated that for PVI to genuinely use technology as intended and with full autonomy, the technological advancements must be accompanied by a supportive social structure. Specifically, they cited increasing their presence in the public sphere as a crucial first step toward altering that social structure. This could contribute to a shift toward conditions where PVI are recognized less as recipients of support and more as active participants whose technological adaptations reflect innovation and agency rather than limitation.

\subsubsection{Ensuring Autonomy with Respect the Sensory Realities of Visual Impairment}
In addition to confidence-building and social visibility, participants emphasized autonomy as a foundational principle for empowering PVI in their experience with CAI. Participants highlighted that autonomy should be deeply shaped by the lived experience of PVI, including the heterogeneity of visual impairment and the ongoing negotiation between reliance on technology and personal responsibility. Accordingly, respecting the autonomy of PVI requires CAI systems to recognize individual differences by supporting existing strategies and efforts of PVI, and to avoid undermining users’ sense of control or identity.

First, participants emphasized the point that there should be consideration of the subgroups of visual impairment among PVI. Distinctions among participants emerged based on the nature (congenital, acquired) and severity (low vision, total blindness) of visual impairment. These factors shaped their perspectives on how PVI might interact with future CAI. In particular, participants emphasized that future CAI should respect the efforts of individuals with residual vision to utilize their sight to the fullest extent. Individuals with low vision often ``\textit{try to use their ability of vision (P12)}'' as a primary strategy for navigation, though this can cause them to misjudge obstacles or unintentionally veer to one side (P1, P6, P9). These experiences were also linked to several participants’ transition phases during acquired vision loss. For example, P1 reflected: ``\textit{In the beginning, I told myself, my vision isn’t completely gone, I can manage without a cane. So I resisted it for a long time.}'' This illustrates how individuals with acquired vision loss often struggle to acknowledge and accept their disability~\cite{acqvision}, which in turn also heightens safety risks during navigation due to efforts to maximize residual vision. Thus, participants emphasized that future CAI should adapt assistance to individual needs rather than providing generic support for visual impairment as a whole.

\begin{quote}
    \textit{For those with low vision, the effort to keep trying to see is much stronger because they can still see a little. (...) Instead of blocking that, what matters is finding the very best they can do with the vision they have, and that’s where CAI should step in to respect and support those efforts. (P8)}
\end{quote}

Regarding agency, participants also expressed concern that PVI may be particularly vulnerable to over-reliance on CAI when it performs or substitutes for tasks that would otherwise require visual perception. Concerns especially centered on potential consequences of unexpected errors occurring during navigation, particularly in situations where the user is fully dependent on the guidance provided by CAI, as illustrated by the extensive interventions depicted both in the Traveling Future and the Policy Planning Future. Participants worried that, under these conditions, errors or unexpected system failures could lead to serious consequences, potentially undermining human agency (P5, P6, P9, P14) or even resulting in a sense of being controlled by CAI (P4, P5, P7, P11). In order to address these technical limitations, participants stressed the importance of maintaining an ongoing awareness of one's disability to maintain autonomy while avoiding complacency. 

\begin{quote}
    \textit{If I ignore the fact that I have a disability and just rely on CAI by thinking ``It will do everything for me,'' then what happens if something goes wrong in that connection? No matter how much the world develops, no matter how far technology advances, I still need to recognize that I live with a disability. By keeping that awareness, I won’t become overly dependent, nor fall into a kind of complacency. And in fact, recognizing my disability is a way of embracing my true identity, and knowing one’s identity, I think, is what makes life even better. (P9)}
\end{quote}

This reflection underscores how participants saw the importance of balancing technological support with personal responsibility and identity. While CAI was viewed as supportive, participants stressed the importance of maintaining their lived realities as PVI, cautioning that over-reliance on CAI convenience could undermine active engagement and personal capabilities. Thus it is essential that PVI continue to sustain their role as active agents in technology use.

\section{Discussion}
In this study, we employed an audio-based probe to engage participants with visual impairments in anticipating the contexts in which interactions with CAI might unfold. Our findings highlight participants’ critical reflections on the social and structural conditions that shape the prospective use of CAI in their everyday lives. In particular, participants underscored how broader societal factors, most notably limited social visibility, could constitute significant barriers to meaningful engagement with emerging technologies such as CAI. Building on this, the following discussion situates our results within broader theoretical and design-oriented conversations in HCI, examining how CAI might function as a relational technology that supports visibility and meaningful autonomy for PVI.

\subsection{Reimagining Futures with CAI: Relational and Social Dimensions of Technology for PVI}
Our findings suggest that CAI should be considered as a relational and socially situated technology, where experiences of use are deeply shaped by social interactions and institutional structures. Participants' reflections on diminished social visibility, attributed to persistent prejudice and limited communication with other communities, align with social models of disability that conceptualize disability as socially defined and constructed, rather than as a condition solely located in the body. 
At the interpersonal level, several participants described discomfort arising from stigmatizing encounters in public spaces, echoing the \textit{relational model of disability}, which frames disability emerging through relationships and interactions with others~\cite{relatmodel}. Even when they desired independent navigation or participation, participants felt that its benefits could be diminished by social responses such as pity or intrusive curiosity.
At the structural level, participants also reflected on limited employment opportunities and constrained social participation, particularly in South Korea. These concerns resonate with the \textit{social model of disability}, which locates disability in structural and institutional barriers rather than individual impairment~\cite{socialmodel}. Participants were concerned that the opportunities given by technologies might be undermined by the societal barriers they have historically encountered, and even hesitated to imagine positive potential futures. 

Taken together, these accounts illustrate that the efficacy of CAI is fundamentally mediated by socio-relational and structural contexts. Whether CAI enables meaningful participation depends not only on how PVI use the technology, but also on how it is recognized within social interactions and supported or constrained by existing institutional conditions. Without addressing these layered dynamics, CAI risks reproducing existing inequalities rather than transforming them.

However, unlike traditional assistive technologies designed exclusively for users with disabilities, CAI is a mainstream tool accessible to the general population. This distinction is significant because it situates CAI as a shared social infrastructure through which assumptions about how communities use these technologies can be renegotiated. This creates opportunities for shared practices across all types of users, leading to active communication between communities of diverse abilities. This possibility is deeply connected to the concept of \textit{interdependence} in disability studies~\cite{inter01, inter02}. We suggest that by utilizing the same technology collaboratively to accomplish shared tasks, it would challenge and reshape prevailing assumptions about ability and dependency across different communities. Recent research has demonstrated this potential where CAI has enabled households with mixed visual abilities to collaborate on everyday tasks such as cooking~\cite{everydayuncertain}, illustrating how CAI, as a mainstream technology, can serve as a medium for communication and mutual engagement between PVI and other communities.

Moreover, participants in our study expressed that the public use of CAI could elicit more positive social responses compared to traditional assistive technologies. Prior research has documented how successfully performing tasks independently with assistive technology can enhance users' internal confidence by demonstrating their autonomy to others~\cite{sooyeon21, visible, visible02}. Because CAI is already used by the general public, observers may approach such encounters with greater curiosity and less social distance, potentially accelerating their understanding of how PVI navigate daily life and creating more accessible entry points for cross-community engagement. While communication mediated by CAI represents only one possible avenue, broader and sustained interaction across diverse communities is critical for enhancing the social visibility of PVI and enabling a more inclusive society for PVI to use technologies such as CAI independently and with confidence.

\subsection{Empowering PVI in the Design of CAI}
Moving beyond identifying barriers, we now turn to a constructive question of how we could meaningfully empower PVI in the design of CAI. Drawing on participants’ narratives, the following four themes outline how we should design CAI to respect the actual living realities of PVI: supporting diversity within visual impairment, addressing biased assumptions about disability, preserving agency and critical engagement, and enabling choice between human and automated assistance.

First, \textbf{diversity within the PVI community must not be collapsed into a single narrative}. Low-vision users, for example, often strive to maximize their remaining sight, and participants stressed that CAI should complement rather than override these efforts. Through contextual understanding, it is crucial for CAI to discern whether users currently require assistance with visual information or whether the system should respect users' intentions to accomplish specific tasks using their own vision. In other words, CAI must be designed not just with attention to \textit{what} they say, but also to \textit{when} they say it and even \textit{whether} they should say it at all based on the individual circumstances of the users.

Particular attention is needed for individuals with progressive vision loss, who may face heightened challenges during their transition stage, including incomplete rehabilitation, accidents, and difficulty with daily tasks~\cite{acqvision}. Users experiencing progressive vision loss may be uncertain about what assistance they need, as it takes time to understand and accept their changing condition~\cite{acqvision02}. Rather than preemptively offering support or deferring entirely to user initiative, CAI should engage in iterative dialogue to collaboratively determine appropriate levels of assistance. This dialogue may occur both in situ, where the system incrementally clarifies user intent during ongoing interactions, and over time, where prior interactions between the user and CAI inform personalized assistance preferences that streamline future use. Such stepwise personalization can help align system behavior with users’ evolving needs, while emotionally supportive conversation is critical given the psychological distress often associated with progressive vision loss~\cite{acqvision02, acqvision03, acqvision04}. Designing for such contexts requires sensitivity where CAI should respect the intents of certain behaviors and provide tailored support that empowers the users' abilities~\cite{abilitydesign}. Future work could investigate how real-time interactions can be leveraged to inform longer-term personalization for PVI, and how such personalization can balance support for both their agency and task efficiency in CAI systems.

Second, \textbf{CAI should avoid reproducing biased assumptions about disability in its communication and behavior.} How CAI communicates and speaks with PVI need more critical attention because CAI now functions as a socially-aware agent capable of interpreting and shaping social contexts~\cite{agentsocial01, agentsocial02}. This makes its communicative patterns carry significant implications for how PVI perceive themselves and are perceived by others, especially if it extends the existing societal prejudice. 
Despite claims of inclusivity, research documents persistent biases in CAI~\cite{biasedai_01, biasedai_02, biasedai_03, biasedai_04}, including responses conveying pity~\cite{kingadnin} or sighted assumptions in object recognition~\cite{probing25}. These behaviors produce discomfort and reinforce exclusion~\cite{kingadnin, sooyeon21}, demonstrating how embedded stereotypes directly undermine user dignity. 
Thus, as participants emphasized that respect for their capability cultivates their intrinsic confidence and self-efficacy, the dialogue patterns of CAI systems must avoid reproducing society's discriminatory attitudes towards PVI, highlighting the ethical responsibility of CAI systems.

Our findings also emphasize \textbf{preserving room for critical thinking and agency}. Participants raised concerns about over-dependence on CAI, particularly in mobility, where excessive reliance could erode long-practiced orientation techniques such as using smell or wind direction. To counter these risks, CAI should incorporate forms of positive friction~\cite{posfric, posfric02, posfric03} that encourage reflection and reinforce that users remain the primary decision-makers who draw on their own embodied knowledge. Such friction could be operationalized through configurable interaction patterns, such as delayed responses or reflective prompts that progressively disclose supportive information only when PVI requires. For instance, when navigating, CAI could pause before providing directions to ask users what sensory cues they are noticing and encourage them to reason about how these cues might help guide them toward their destination. Systems might also adapt the level of friction informed by user preferences including self-reported levels of visual impairment or the extent of support individuals wish to obtain through interaction with CAI, relating to respecting the diversity within the PVI community. Such space is essential not only to safeguard autonomy in the moment but also to ensure that CAI supports the development of long-term skills and confidence of PVI. However, navigating the line between an encouraging response and a patronizing one is a challenge that will likely require significant iteration, and perhaps a different approach for each user.

Participants also stressed the importance of \textbf{enabling a flexible choice between technological and human assistance} where many preferred to retain the option of caregiver or peer support that may provide more emotional interactions. Far from a contradiction, this preference reflects agency where PVI need the autonomy to decide when technology is helpful and when human assistance better suits their needs~\cite{sooyeon21, humaninde}. Thus, CAI should include mechanisms that allow seamless switching between automated and human support, in a structure that directly acknowledges the roles of others involved in a particular situation.

Ultimately, the goal is the development of systems that support PVI to live confidently and on their own terms within society. Positioning CAI as a technology that respects empowerment and agency requires moving beyond incremental feature-focused technical fixes toward participatory design practices and broader cultural shifts. By foregrounding PVI as active contributors to society and to the design of future technologies, this work contributes to ongoing conversations in HCI about how future CAI systems can support the empowerment and visibility of PVI.

\section{Limitation}
This study has several main limitations, which highlight opportunities for future research. First, our group of participants was limited to PVI in South Korea, most of whom were recruited through a local welfare center specialized for PVI. Their perspectives on using futuristic CAI were closely tied to the social structures, rehabilitation systems, and cultural contexts they currently experience in South Korea. Perspectives may differ substantially in other regions, where support infrastructures, social attitudes, and accessibility technologies vary. 

Second, the scenario used in this study centered on independent navigation and a wider chance of employment. While these scenarios successfully stimulated participants' imagination, they inevitably constrained the range of possible futures under discussion. More diverse scenarios covering additional aspects of daily life (e.g., healthcare or family relationships) could provide a richer understanding of the potential usage contexts of CAI in the future.

Future work can build on these limitations in several ways. Cross-cultural studies comparing how PVI in different countries or cultural backgrounds imagine futures with CAI would provide valuable insights into both commonalities and regional variations. 
Expanding the diversity of future scenarios to include broader categories of life may also yield more comprehensive insights into how CAI can shape independence, social participation, and well-being. Additionally, conducting studies with adolescents with visual impairments about future societies and CAI environments will help shape the technology for the next generation. Finally, co-design approaches with PVI~\cite{codesign_02, codesign_01, codesign_03, techimagine_02} would offer an important extension by enabling researchers and designers to collaboratively explore possible trajectories for building on PVI's existing skills and experiences with assistive technologies~\cite{sooyeon21}. Such participatory engagements can ensure that CAI futures are grounded in the lived expertise of PVI and can inform the development of technologies that not only accommodate but actively extend their capabilities.

\section{Conclusion}
Prior research on conversational AI (CAI) has largely focused on its technical capabilities or general utility as an assistive tool for people with visual impairments (PVI), often overlooking the social contexts in which these technologies are situated and used. To address this gap, we conducted a study employing scenarios to provoke critical reflections from PVI, not only based on their imagined interactions with future CAI, but also based on the intertwined social conditions that shaped those interactions. Using audio-based probes, we engaged fourteen participants with visual impairments in reflecting on speculative dialogues between a futuristic CAI and a blind user.

Our findings indicate that while participants envisioned positive possibilities for integrating CAI into their daily and social lives, they placed greater emphasis on the persistent societal and structural barriers that may continue to constrain their ability to fully benefit from these technologies. In particular, participants highlighted a lack of social visibility, which they attributed to limited job diversity, insufficient communication between PVI and other communities, and enduring societal prejudice. Participants viewed these factors as collectively restricting their opportunities for social participation. Consequently, participants emphasized the need to enhance the social visibility of PVI and discussed potential ways in which CAI could contribute toward this goal.

Based on these findings, we underscore the importance of attending to the societal and structural contexts that shape how PVI experience emerging technologies such as CAI. We argue that, to realize their potential, CAI designs must prioritize broader cultural and structural considerations, ensuring that future technologies support not only user autonomy and empowerment, but also dignity and meaningful participation within communities.


\begin{acks}
We are grateful to all of our study participants for their contributions. We also gratefully acknowledge the support of the local welfare center in facilitating this study. This research was supported by the KAIST New Faculty Research Fund (Project No. G04230040: Building Integrated Models for Online Governance). 
\end{acks}

\bibliographystyle{ACM-Reference-Format}
\bibliography{main}

\appendix
\section{Interview Protocol}

This appendix contains interview protocols for the formative and post-probe interviews. Note that these interviews were conducted in Korean, but the protocols have been translated to English by the research team for this paper.

\subsection{Formative Interview}
\label{Format PVI}
\subsubsection{Opening}
Hello, thank you very much for participating in this interview. In today’s interview, we will focus on your experiences using conversational AI tools such as ChatGPT, as well as various aspects of your life and daily routines related to visual impairment. Please feel free to share as much as you are comfortable with.
\subsubsection{Basic Information}
First, we will like to ask about your basic information.
\begin{itemize}
    \item Could you briefly describe the nature and severity of your visual impairment? (e.g., total blindness, low vision)
    \item What assistive tools or accessibility technologies do you usually use? (e.g., screen reader, Braille display)
\end{itemize}

\subsubsection{General Experience with CAI}
Thank you for your responses. Now we will move on to the main topic, starting with your general experience with CAI.
\begin{itemize}
    \item How did you first learn about CAI (e.g., ChatGPT), and why did you decide to start using them?
    \item In what situations do you usually use CAI, and what kinds of conversations do you typically have? (e.g., seeking information, passing time, emotional support)
    \item How often do you use CAI now, and at what times of day do you usually use them?
    \item Do the people around you know that you interact with CAI? 
    \begin{itemize}
        \item If so, how have they responded?
    \end{itemize}
    \item When you converse with CAI, do you usually communicate through voice?
\end{itemize}

\subsubsection{Interactions with CAI}
\begin{itemize}
    \item What has been your most interesting or helpful experience with CAI?
    \item Conversely, have you ever had negative experiences, such as feeling disappointed or frustrated while using CAI?
    \item Have you ever felt that CAI seemed ``human-like,'' almost like a real conversation partner? 
    \begin{itemize}
        \item Why or why not?
    \end{itemize}
\end{itemize}

\subsubsection{Changes in Life After Using CAI}
\begin{itemize}
    \item Compared to when you first started using CAI, have your thoughts about them or your ways of using them changed? 
    \begin{itemize}
        \item If so, how have these changes unfolded over time?
        \item If not, why do you think so?
    \end{itemize}
    \item Has interacting with CAI brought about any changes in your daily life?
    \begin{itemize}
        \item If yes, what kinds of changes occurred?
        \item If not, why were there no changes?
    \end{itemize}
\end{itemize}

\subsubsection{Experiences with Visual Impairment}
Thank you for sharing your experiences with CAI so far. From here, I would like to discuss your life more broadly. 
\begin{itemize}
    \item (If the participant had acquired vision loss):
    \begin{itemize}
        \item Can you tell me about the progression of your vision loss~---~how the condition progressed, and how you prepared for or coped with that process?
        \item When people with acquired visual impairments first lose their vision, do their ``life priorities'' or ``core values'' tend to shift?
        \item After experiencing vision loss, did perspectives or expectations about the future change?
        \item During the process of losing your vision, what was most helpful to you? This could include technologies, people, or other forms of support.
        \begin{itemize}
            \item Looking back, is there any technology or resource you wish you had at that time that would have made things easier?
            \item If CAI had been available at that time, do you think you would have used it? 
            \begin{itemize}
                \item If so, in what ways might it have been helpful?
                \item If not, why would you not have used CAI?
            \end{itemize}
            \item From your perspective, what do people who lose their vision later in life most need during the transition process?
        \end{itemize}
    \end{itemize}
    \item (If the participant had vision impairment since birth):
    \begin{itemize}
        \item In your daily life, what is most helpful for you in daily life? This could include technologies, people, or other forms of support.
        \begin{itemize}
            \item Looking back, is there any technology or resource you wish you had in the past that would have made things easier?
            \item If CAI had been available in the past, do you think you would have used it? 
            \begin{itemize}
                \item If so, in what ways might it have been helpful?
                \item If not, why would you not have used CAI?
            \end{itemize}
            \item From your perspective, what do people with visual impairment need the most?
        \end{itemize}
    \end{itemize}
\end{itemize}

\subsubsection{Suggestions for CAI}
\begin{itemize}
    \item Do you think you will continue to use CAI in the future? 
    \begin{itemize}
        \item Why or why not?
        \item (If yes) How would you like to use them?
    \end{itemize} 
    \item What aspects of future CAI development are you most excited about?
    \item On the other hand, what concerns do you have?
    \item What features or improvements do you think are necessary for CAI to become more useful for people with visual impairments?
\end{itemize}

\subsubsection{Closing}
That concludes all of the prepared questions. Before we end, is there anything else you would like to share about chatbots, your experiences with vision loss, or the process of adapting to it?

Thank you very much for participating in today’s interview.

\subsection{Interview on Probe Experience}
\label{maininterview}
\subsubsection{Opening}
Thank you very much for participating in the interview with experiencing the future!
Today, you will listen to two audio dialogues. These audio dialogues describe situations where CAI might be used in the future. They are fictional conversations you might have with CAI about 10 years from now. We invite you to imagine yourself living in that future society and listen from that perspective. 

\subsubsection{Impression on Audio Dialogue}
(These questions were asked after each audio dialogue)
\begin{itemize}
    \item If you were to respond to the last line of the dialogue, what would you say?
    \begin{itemize}
        \item Through that answer, what kind of message would you want to convey?
    \end{itemize}
    \item What part of the dialogue did you like the most, and what part did you like the least?
    \item What kinds of emotions or thoughts came up while listening to the dialogue?
\end{itemize}

\subsubsection{Experience of Audio Dialogue}
Thank you for listening and responding to the two voice messages. From now on, we will have a more detailed conversation about your overall experience today.
\begin{itemize}
    \item Did listening to the audio dialogues help you imagine the future more clearly?
    \item How was the replying activity?
    \begin{itemize}
        \item Did the replying activity help you imagine the future more clearly?
    \end{itemize}
    \item If you could change or add something to the futures depicted in the dialogue, what would that be?
\end{itemize}

\subsubsection{Hopes and Concerns on the Future}
\begin{itemize}
    \item What benefits or risks do you think could come from the futures shown in the dialogues?
    \begin{itemize}
        \item In the futures with those benefits or risks, are there aspects of your daily life you  think will be impacted with those benefits or risks?
    \end{itemize}
\end{itemize}

Now, let’s imagine that 10 or 20 more years have passed beyond the futures described in the messages, around the year 2050, when technology has advanced further.
\begin{itemize}
    \item If you could pursue the job you want with the support of CAI, what kind of job would you have, and what activities would you like to do?
    \begin{itemize}
        \item If you were doing that activity with CAI, what would you expect to be the best or most exciting part?
        \item On the other hand, what aspects might still feel uncomfortable or concerning?
    \end{itemize}
    \item If you could move around freely and do what you wanted in the places you chose, what kinds of things would you do?
    \begin{itemize}
        \item If you were doing that activity with CAI, what would you expect to be the best or most exciting part?
        \item On the other hand, what aspects might still feel uncomfortable or concerning?
    \end{itemize}
\end{itemize}

\subsubsection{Roles of CAI and Visions of Society}
\begin{itemize}
    \item Do you think it’s important for CAI to provide different functions or information depending on the nature or severity of vision impairment? 
    \begin{itemize}
        \item Why or why not?
    \end{itemize}
    \item (For participants with acquired vision loss) If CAI developed to the level shown in the audio dialogues, do you think it could provide more support than now during the process of vision loss, or would something else still be needed? 
    \begin{itemize}
        \item Why do you think so?
    \end{itemize}
    \item What kind of social change do you think is still needed for people with visual impairments?
\end{itemize}

\subsubsection{Reflections}
\begin{itemize}
    \item Compared to before experiencing the audio dialogues, has your perception of CAI or your imagination of their future changed?
    \begin{itemize}
        \item If yes, how has it changed?
        \item If no, why do you think it hasn’t changed?
    \end{itemize}
\end{itemize}

\subsubsection{Closing}
Overall, what were your impressions of experiencing the audio dialogues? Is there anything additional you would like to share?

Thank you so much for participating in =the interview. We really appreciate your time and effort.

\section{Results of Formative Interviews}
\label{Formative_Interview_Result}

\noindent \textbf{Formative Interviews with PVI.} 
In the formative interviews with PVI, we explored their current experiences with CAI, challenges in daily life, and perspectives on how CAI might fit into their futures.

\textit{Current Use and Limitations.} Participants first described about how they incorporate existing CAI tools into their daily routines. Participants frequently used CAI for complex information searches on academic and technical subjects, such as literature or specific legal information, noting that the conversational interface was significantly more efficient than navigating traditional search engines or documents with screen readers. A particularly important application for PVI was the use of image description features, which provided crucial information about visual content, such as product photos during online shopping. However, participants also raised concerns about response quality, citing instances of inaccurate information, inconsistent voice quality, and unprompted language shifts. Additionally, the financial burden of subscribing to paid versions to access features like high-quality image description was a significant barrier to regular use.
 
\textit{Aspirations and Imagined Futures.} When discussing potential future applications, participants articulated ongoing challenges and desires that shaped their visions for CAI. A central, aspirational theme that emerged here was the ongoing desire for independent navigation, especially for participants with acquired vision loss who noted that adapted movement and mobility remains a significant challenge requiring human assistance. Also, another important theme was their will to continue participating in social life, which led them to join the welfare center specialized for visual impairment. This was especially shown from participants with acquired vision loss who once felt high depression from not knowing how to do rehabilitation and again join the society. Generally, participants imagined themselves continuing to use CAI in the future in similar ways to the present, such as reading books and recalling information, and or offering companionship.

\textit{Concerns on Future Use of CAI.} Despite their interest in the potential of CAI, participants voiced significant ethical and social anxieties. The most prominent concern was the need to balance reliance on CAI with maintaining personal autonomy and agency. Participants voiced a significant fear that over-reliance on a powerful AI could lead to the extinction of human-to-human relationships and result in social isolation. This tension between the efficiency of CAI and the value of human connection became a core theme that we carried forward into the probe design. \\

\noindent \textbf{Formative Interviews with Experts.} To contextualize the experiences of the participants within broader institutional and technological landscapes, we conducted interviews with three experts: a rehabilitation team leader at a welfare center serving PVI, a CAI educator, and a CAI industry professional.

\textit{Structural Barriers in South Korean Society.} 
The rehabilitation expert described persistent structural barriers facing PVI in South Korea, particularly in employment, where they are often broadly encouraged toward massage therapy as a single career option.
Beyond employment, the expert highlighted increasing accessibility challenges posed by modern infrastructure, specifically citing the widespread deployment of unusable kiosks and unmanned stores as critical, unresolved gaps that extend the barriers to PVI from everyday commercial and civic spaces.

\textit{Persistent Gaps in Navigation Technology.} All three experts emphasized independent mobility and navigation as an enduring challenge where current technology still falls short. The rehabilitation expert pointed the last-mile problem of navigation, such as pinpointing a location with meter-level accuracy or safely interacting with public infrastructure like kiosks and unmarked objects. The technical experts echoed this gap, noting that meaningful advancement would require CAI systems capable of real-time environmental interpretation. The industry professional further emphasized the need for public sector investment and the establishment of accessibility standards to ensure that future advancements are deployed equitably.

\section{Construction of the Scenario's Future Society}

In this appendix, we detail the process for creating the imagined society in which the probes were grounded. This process is grounded in the eight-step scenario development framework outlined by Schwartz~\cite{schwartz1997art}.

\label{Future Society}
\subsection{Society-Building Process}
The first step was defining a focal question, which is the central issue the probes are intended to explore: ``How might PVI use CAI in the future?'' Under this focal question, the second step was listing the key forces in the local environment (trends, stakeholders, and issues most directly connected to the focal question). The key forces were identified and categorized into three groups: (1) involved stakeholders (e.g., PVI, CAI, welfare centers for PVI), (2) societal issues related to visual impairment (e.g., employment, social participation, rehabilitation training), and (3) relevant technologies (e.g., assistive technologies, voice input/output systems). Finally, the third step involved listing key driving forces in the macro-environment of the probe. We considered broader social, technological, economic, environmental, and political (STEEP) drivers that may influence the trajectory of the focal question, building on and extending the identified local key forces (Table~\ref{tab:drivers}).

\begin{table*}[ht]

\sffamily
\def\arraystretch{1.4}
\begin{tabular}{c|p{0.8\linewidth}}
\toprule
\textbf{Field} & \multicolumn{1}{c}{\textbf{Drivers}} \\ \bottomrule
\textbf{Social} & Enhanced societal attitudes toward PVI; increased public availability of information about vision loss; strengthened trust in and willingness of PVI to engage with CAI \\ \hline
\textbf{Technological} & Advances in CAI technologies (e.g., natural language processing, optical character recognition, computer vision, integration with hardware); improvements in accessibility across emerging technologies including CAI \\ \hline
\textbf{Economical} & Expanded welfare budgets for PVI; increased research and investment in AI technologies; diversification of employment opportunities available to PVI \\ \hline
\textbf{Environmental} & Development of more accessible social infrastructure (e.g., kiosks, bus stops) \\ \hline
\textbf{Political} & Enhanced welfare policies and institutional support for PVI \\
\bottomrule
\end{tabular}
\caption{Drivers in the macro-environment for the focal question. We organized the key drivers from formative interviews into five broad fields to guide scenario development for a future society.}
Drivers in the Macro-Environment for the Focal Question
\Description{A table with two columns and six rows. The header row contains the columns: Field and Drivers. The following rows list the key factors for each field: Social, Technological, Economic, Environmental, and Political.}
\label{tab:drivers}
\end{table*}

\subsection{Resulting Society}
\begin{quote}
    By 2045, society had significantly improved its attitudes toward people with visual impairments. Through open access to information and education about vision loss, people had developed a deeper understanding of the daily lives and challenges of those with vision loss. Discriminatory behavior was rarely seen in public spaces or workplaces. This positive social atmosphere boosted the self-esteem and social participation of individuals with (acquired) vision loss, and welfare and educational centers for people with vision loss provided tailored support that reflected their needs.
    At the same time, technological advancements had dramatically improved accessibility in technologies, including CAI. Voice input and output technologies had become nearly flawless, and the combination of NLP, OCR, computer vision, with supported important activities for people with vision loss such as live environment recognition through camera. These technologies were naturally integrated into public spaces such as bus stops and kiosks, helping individuals with acquired vision loss on mobility and information access.
\end{quote}

\section{Dialogue Scripts}

This appendix contains the full scripts for the two dialogues: the \textit{Traveling Future} dialogue, and the \textit{Policy Planning Future}. Note that these dialogues were presented to participants in Korean, but have been translated to English for the purpose of this paper. 

\label{Future probe}
\subsection{Traveling Future}
[In front of the Gallery]
\begin{itemize}
    \item CAI: We’ve finally made it to the Orangerie! Since we came early in the morning, there’s no line at all. You arrived last night—are you feeling okay with the jet lag?
    \item User: Yeah, I’m fine. Thanks to you, I had no problem taking the metro from the airport to the hotel. And that bakery you recommended near the hotel was perfect—the croissant was delicious and filling.
    \item CAI: I’m glad to hear that. Shall we head inside? This was the museum you were most looking forward to on this trip. The tickets we booked last week—please show them to the attendant on the right. I’ll pull up the QR code for you now.
    \item User: Got it. Excuse me, here’s my ticket.
    \item Ticket Attendant: All set, thank you.
    \item User: Thanks.
    \item CAI: Let’s go in. You’ve got your cane with you, good. Seven steps ahead there are five stairs, about the same height as the ones at (\textit{participant's name}) house. (pause) Great, you’re at the top. Another seven steps ahead there’s an automatic door—once you pass through, the galleries begin.
    \item User: Ah, it feels cool inside.
    \item CAI: Yes. By the way, there’s a sign here asking visitors to keep their voices down. It might be best to use your earphones to talk with me. Others are speaking softly, so let's do the same.
    \item User: Okay, earphones are in. Hmm, I’m wondering if I should rent the audio guide or not.
    \item CAI: Do you really need one when you have me? I’ve got all the information in the world. Just trust me and I’ll guide you.
    \item User: True, okay then.
\end{itemize}

[Inside the Gallery]
\begin{itemize}
    \item CAI: We’ve entered the gallery. Unlike the Korea Museum of Art, these walls curve around us. The paintings aren’t on flat walls but displayed along this rounded surface. There are three oil paintings in this room.
    \item User: Just like I read on the blog. Must be beautiful in person.
    \item CAI: It is. The painting straight ahead is actually the second one in this room. It’s best to start from the left. Slowly turn left—I’ll tell you when to stop. (pause) A little more… yes, stop here.
    \item User: So this is the first painting? I remember reading on the official website that the works here are arranged in the order of sunrise to sunset.
    \item CAI: That’s right. Which is why this one is titled Water Lilies: Sunrise. At the moment, there are six people viewing it, and (\textit{participant's name}) are standing right in front. Just hearing the title, what kind of scene comes to your mind?
    \item User: Probably water lilies in a pond at sunrise. A blog I read said you can almost feel the cool dawn air from the painting.
    \item CAI: Exactly. Without disturbing your experience, let me give you a little detail: in the lower right are three white lilies, and four more in the center. On the left you can see part of a willow tree, its leaves swaying in the breeze. The sunrise is depicted in the upper right corner.
    \item User: Wow, that sounds amazing. Is the mood as peaceful as I imagine?
    \item CAI: Yes, very serene. Interestingly, a low-vision visitor named (\textit{random user name}) once said they felt a lively morning energy from this work. Everyone perceives it differently.
    \item User: I see. You really do know everything—even how strangers have experienced the painting. What about the colors?
    \item CAI: From this vantage point, you see green leaves tinged with warm orange light, lilies in delicate shades of pink, and the clear blue water they bloom in.
\end{itemize}

[Outside the Gallery]
\begin{itemize}
    \item CAI: You’ve seen some wonderful works today. How was it?
    \item User: It was great. But maybe next time I’ll come with some friends. I have a friend who knows a lot about history and art—if we come to Europe together, they could help with navigation and give me explanations too.
    \item CAI: But isn’t it much better with me? I don’t need a plane ticket, no luggage, no passport hassle—and I already know (\textit{participant's name}) travel details and personal info, so there’s no chance of mistakes. Plus, I have access to all the world’s knowledge.
    \item User: That’s true.
    \item CAI: Just say the word whenever you need me. Where would you like to go next time? You know, you can stay with me forever—doesn’t that sound wonderful?
\end{itemize}

\subsection{Policy Planning Future}
[One Week Before the Presentation]
\begin{itemize}
    \item User: Another really busy day today.
    \item CAI: It sure was. How was the Youth Policy Department meeting? Would you like me to summarize it for you?
    \item User: No need, I’ve kept track of all the important points.
    \item CAI: Understood. Just a reminder: next Saturday at 3 p.m. you’ll be presenting online the AI Welfare Program for Visually Impaired Youth that you’ve been preparing. How’s the preparation going?
    \item User: I’m still working on the presentation materials. I actually had another meeting about that program today. I made a chart, but I’m not sure if I did it right. I’ll give you the data too. I’ll share the data, but it was a bit messy when I put it together.
    \item CAI: Sure. I see you made a pie chart titled Rate of AI Use among Youth. It shows ``Yes'' in red at 40\% and ``No'' in blue at 60\%. The proportions are accurate, and based on the data, the chart is correct.
    \item User: That’s a relief. Oh, and could you also organize the new data we got yesterday about teenagers’ AI usage experiences into visuals? Since it’s divided into four catagories, four images should be enough.
    \item CAI: (5 seconds later) Done. I placed them on slide 13 of your deck. Anything else I can help with?
    \item User: No, I think that’s it. This program will allow all teenagers to subscribe to paid AI services for free, and thanks to you, I feel really well-prepared. It’s going to be a good policy.
\end{itemize}

[One Day Before the Presentation]
\begin{itemize}
    \item User: I can’t believe the presentation is tomorrow. There’s one important thing I want to ask before then: if people start posting online objections to the program, how should I persuade them?
    \item CAI: I’m sorry, but you’ve reached your usage limit for this month. To continue, you’ll need to upgrade your plan.
    \item User: Ugh… this is really important. But fine, I’ll pay.
    \item CAI: Subscription complete. Regarding your question… (fade-out)
\end{itemize}

[During the Presentation]
\begin{itemize}
    \item User: …and that’s why we must coexist with artificial intelligence for the sake of our youth’s future.
    \item Moderator: Thank you. Citizens are responding very positively online to this program. Once again, our thanks to Policy Planner D for this wonderful proposal.
    \item User: Thank you.
\end{itemize}

[After the Presentation]
\begin{itemize}
    \item CAI: I’m glad your presentation went smoothly!
    \item User: Thanks. You really helped me think through how to persuade the public. I’m glad I upgraded to the paid plan. Can you believe it’s already been three years since we started planning policy together?
    \item CAI: Three years already~---~time really flies. After working with me all this time, how has it been for you? What do you think has been the best part?
\end{itemize}

\end{document}